\documentclass[aps,prb,twocolumn,showpacs]{revtex4}
\usepackage{graphicx}
\usepackage{textcomp}

\begin{document}

\title{Carbon antisite clusters in SiC: a possible pathway to the
  D$_{\text{II}}$ center}

\author{Alexander Mattausch}
\email{Mattausch@physik.uni-erlangen.de}

\author{Michel Bockstedte}

\author{Oleg Pankratov}

\affiliation{Lst. f\"ur theoretische Festk\"orperphysik, Universit\"at
  Erlangen-N\"urnberg, Staudtstr. 7, D-91058 Erlangen, Germany}
\date{\today}

\begin{abstract}
  The photoluminescence center D$_{\text{II}}$ is a persistent
  intrinsic defect which is common in all SiC polytypes. Its
  fingerprints are the characteristic phonon replicas in luminescence
  spectra. We perform ab-initio calculations of vibrational spectra
  for various defect complexes and find that carbon antisite clusters
  exhibit vibrational modes in the frequency range of the
  D$_{\text{II}}$ spectrum. The clusters possess very high binding
  energies which guarantee their thermal stability---a known feature
  of the D$_{\text{II}}$ center. The dicarbon antisite
  (C$_{2}$)$_{\text{Si}}$ (two carbon atoms sharing a silicon site) is
  an important building block of these clusters.
\end{abstract}


\bibliographystyle{apsrev}

\maketitle

\section{Introduction}
\label{sec:introduction}

Many unique features of SiC, such as the wide band gap and the very
high electrical and thermal stability, render this semiconductor
especially important for high power, high frequency and high
temperature applications. As for any semiconductor material, the
identification and understanding of structural defects and impurities
is the key to the technological control of SiC. This is especially
important in connection with the ion-implantation, which is inevitably
accompanied by the generation of intrinsic defects. Some of these
defects are thermally very stable. Two important examples are the
photoluminescence (PL) centers D$_{\text{I}}$ (Ref.~\onlinecite{Pa72})
and D$_{\text{II}}$ (Ref.~\onlinecite{Pa73}).  Although the centers
have been extensively studied over the past 30
years,\cite{Pa72,Pa73,EgBeIv99,Fr87,Sr98_d2,Sr01,Ca03} their
microscopic origin remained unclear. In this work we study carbon
antisite clusters and their possible relevance to D$_{\text{II}}$-type
centers.

D$_{\text{II}}$ centers are always present in ion-implanted and
annealed SiC samples, regardless of the implanted species and the
polytype. They were also found (in low concentration) in as-grown
material.\cite{Fr87} The abundance of D$_{\text{II}}$ centers
increases significantly at high annealing temperatures (above
1300\textcelsius) and the center is thermally stable up to
1700\textcelsius.\cite{Fr87} These properties imply that the
D$_{\text{II}}$ centers are intrinsic defect complexes.  The
D$_{\text{II}}$ luminescence originates from the recombination of an
exciton bound to the defect. The spectra are usually identified by the
main zero-phonon line. The important specific feature of
D$_{\text{II}}$ are characteristic phonon replicas above the SiC
phonon spectrum. In the original experiments in 3C-SiC the strongest
five replicas were highlighted.\cite{Pa73} In later
experiments\cite{Sr98_d2,Ca03} in 4H and 6H-SiC additional localized
vibrational modes (LVMs) were observed (more than 12 LVMs have been
counted in 4H-SiC\cite{Ca03}). Many of these observed modes are
polytype-independent. A comparison of the spectra from
Ref.~\onlinecite{Sr98_d2} and the earlier Ref.~\onlinecite{Pa73}
reveals that more than the five highlighted LVMs may also be contained
in the 3C spectrum. Their spectral density resembles the phonon
density of states in diamond, indicating that a carbon-dominated
defect is responsible for the observed spectrum.

\begin{figure}
  \includegraphics[width=0.7\linewidth]{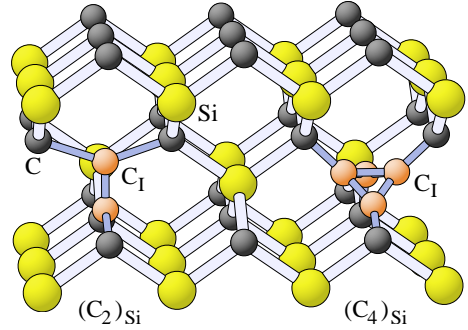}
  \caption{\label{fig:structure} Structure of the dicarbon antisite
    (C$_{2}$)$_{\text{Si}}$ and a carbon cluster with four atoms on a
    silicon site (C$_{4}$)$_{\text{Si}}$ in 3C-SiC.}
\end{figure}

In this work we consider a number of carbon-related defects in SiC:
the carbon di-interstitial, which was the first suggested model for
the D$_{\text{II}}$ center,\cite{Pa73} the carbon antisite, the carbon
split-interstitials and several carbon clusters.  Comparing the
calculated LVMs with the experimental D$_{\text{II}}$ spectra we are
able to unequivocally rule out all candidates except the carbon
antisite clusters. The latter are the only defects that provide a rich
LVM spectrum similar to that of the D$_{\text{II}}$ center. The core
structure of all these clusters is the dicarbon antisite
(C$_{2}$)$_{\text{Si}}$ (cf.  Fig.~\ref{fig:structure} left) which
acts as an aggegation center for further carbon atoms. In this paper
we will first present our results for the dicarbon antisite and then
consider the formation of more complex defect clusters. We calculate
the LVMs associated with these defects and compare them with the
D$_{\text{II}}$ spectrum. Although theoretically all these defects can
acquire different charge states, we focus on the neutral state as the
most relevant for PL. For excitons bound to a charged defect more
non-radiative recombination channels are open, and they are unlikely
to be seen in PL-experiments.\cite{St75} We also find that the
formation kinetics of the carbon clusters agrees with the known
features of the D$_{\text{II}}$ center. These results indicate that
although neither of the considered defects can alone explain all
features of the D$_{\text{II}}$ center, the carbon antisite clusters
may serve as a core structure of D$_{\text{II}}$-type defects.

\section{Method}
\label{sec:method}

We employ an \emph{ab initio} density functional theory (DFT) approach
as implemented in the software package FHI96SPIN.\cite{Bo97} Smooth
norm-conserving pseudopotentials of the Troullier-Martin
type\cite{Fu99} and a plane-wave basis-set with a cut-off energy of
30\,Ry are used. The exchange-correlation potential is approximated
within the LSDA in the parametrization of Perdew and
Zunger.\cite{Pe81} In order to reduce the artificial defect-defect
interaction, large supercells with 216 lattice sites for 3C-SiC and
128 sites for 4H-SiC are used for all calculations of the defect
energetics. For the 216-site cell, the Brillouin zone is sampled by
the $\Gamma$ point, whereas for the hexagonal 128-site cell a special
Monkhorst and Pack\cite{Mo76} $2\times2\times2$ mesh is used. The values of the
formation energies are given assuming silicon-rich conditions. Under
carbon rich conditions they should be reduced by roughly $3
\Delta\/H_{\text{f}} = 1.74$\,eV, where $-\Delta\/H_{\text{f}}$ is the heat of
formation of SiC.  The vibrational properties are calculated using the
\emph{frozen phonon} method. The dynamical matrix
\begin{displaymath}
  \Phi_{ij} = \frac{1}{\sqrt{m_{i} m_{j}}}
  \frac{\partial^{2}E}{\partial Q_{i} Q_{j}} = \frac{1}{\sqrt{m_{i}
      m_{j}}} \frac{\partial F_{i}}{\partial Q_{j}}
\end{displaymath}
is calculated by applying displacements $Q_{j}$ from the equilibrium
configuration and evaluating the force derivatives $\partial F_{i} / \partial
Q_{j}$. To obtain noticeable energy changes, the displacements $Q_{j}$
ought to be much larger than realistic phonon amplitudes.  Therefore
it is important to eliminate the anharmonic contribution when
evaluating $\Phi_{ij}$. We do this by applying three different values of
the displacement and extracting the linear force constant from a
polynomial expansion of the forces. The forces have been converged to
a relative accuracy of $10^{-4}$. For 3C-SiC, the dynamical matrix is
evaluated for a supercell with 64 or 216 sites.  The experimental bulk
phonon modes are reproduced within 5 percent by this method. Yet the
polarization splitting of the TO and the LO modes at the $\Gamma$ point is
missing, since the macroscopic polarization of the crystal is
incompatible with the periodic boundary conditions imposed on the
supercell. We obtain a triple degenerate mode at 115.3\,meV, in place
of the TO mode of 98.7\,meV and the LO mode of 120\,meV.\cite{WiRu99}
For 4H-SiC and for large clusters in 3C-SiC, the computational cost of
evaluating the dynamical matrix of the whole 128-site or 216-site
supercell is prohibitively large. In these cases we constrain the LVM
calculation to the defect molecule (i.e. the dumbbell and its nearest
neighbors) embedded in a supercell. We have examined the error
resulting from the defect molecule approximation and from the
uncertainty in the lattice constant. The results of these tests for a
dicarbon antisite in 3C-SiC (see below) are shown in
Fig.~\ref{fig:modes}. The lines represent the frequencies of the LVMs
versus the defect molecule size calculated at the theoretical (LSDA)
lattice constant. The frequencies shift within the grey area when the
lattice constant is increased towards the experimental value of 8.239
Bohr. The black bars are the same frequency ranges, but in this case
both the electronic structure calculation and the LVM calculation have
been performed in a 64 sites cell. We have also verified that under
the pressure exerted by the periodic array of defects onto the lattice
 the lattice constant lies between the indicated limits. The LVM's
variation due to the uncertainty in the lattice constant is thus less
than 10\,meV.  It is also apparent that the defect molecule
approximation affects the least localized modes 1 and (to a much
lesser extend) 2. The largest effect on the accuracy has the inclusion
of the full supercell, which shifts the mode 1 by 14\,meV in the 216
sites cell. Compared to the full 64 sites cell, the mode 1 is only
2\,meV higher.

\begin{figure}
  \includegraphics[width=\linewidth]{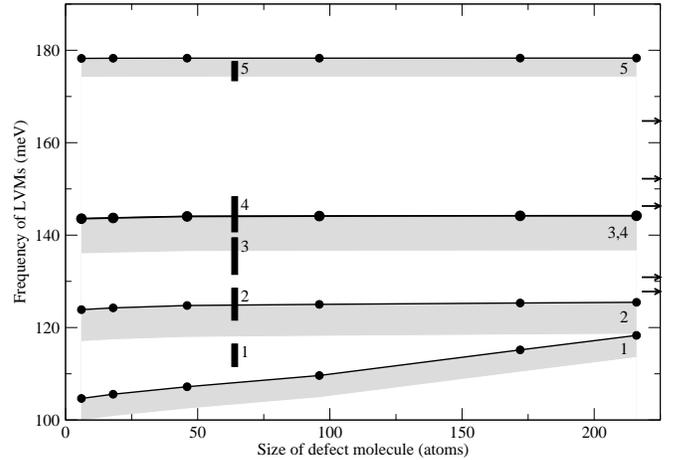}
  \caption{\label{fig:modes} Defect modes of (C$_{2}$)$_{\text{Si}}$
    vs.\ the defect molecule size in a 216 sites cell. The straight
    line denotes the frequency at the theoretical lattice constant,
    the gray area is the frequency shift for the variation of the
    lattice constant between its theoretical and experimental value.
    The black bars are the corresponding values for the 64 sites cell
    as given in Tab.~\ref{tab:LVMs}, the arrows indicate the phonon
    replicas of the D$_{\text{II}}$ center.\cite{Pa73}}
\end{figure}

\section{The dicarbon antisite}
\label{sec:c2si}

Let us now consider one by one the candidates for the D$_{\text{II}}$
center as listed above. First, we find that the originally suggested
carbon di-interstitial has a formation energy of about 12\,eV.  This
is extremely high even for SiC and practically rules out this model. A
similar defect is the carbon split-interstitial C$_{\text{sp} \left<
    100 \right>}$, i.e.\ a pair of carbon atoms sharing the same
(carbon) site and pointing in $\left< 100 \right>$ direction (in
3C-SiC). This defect also possesses a short carbon-carbon bond and it
is energetically the most favorable carbon interstitial.\cite{Ma01}
The second lowest carbon interstitial is the carbon-silicon
split-interstitial C$_{\text{sp\/Si} \left< 100 \right>}$ (a carbon
and a silicon atom on a silicon site). Here the additional carbon atom
is also connected to its nearest neighbors by short carbon-carbon
bonds.  However, according to our calculations both defects are very
mobile,\cite{Bo03b} which is in conflict with the thermal stability of
the D$_{\text{II}}$ center. In addition, we find the LVMs of both
defects are not compatible with the D$_{\text{II}}$ spectrum. The
neutral carbon split-interstitial C$_{\text{sp} \left< 100 \right>}$
in the tilted structure\cite{Bo03b} has two LVMs at 120.4 and
188.7\,meV and LVMs in the phonon gap. The carbon-silicon
split-interstitial C$_{\text{sp\/Si} \left< 100 \right>}^{2+}$
possesses a richer vibrational spectrum but still with only three
non-degenerate LVMs above the bulk spectrum. This also does not match
the number of the characteristic D$_{\text{II}}$ lines.  Another
possible candidate could be the carbon antisite C$_{\text{Si}}$.
Following the carbon vacancy V$_{\text{C}}$ it is the second most
abundant defect in SiC.  However, it shows only vibrational resonances
in the bulk phonon spectrum and has no true LVMs. Also small clusters
of carbon interstitials do not provide LVMs compatible with the
D$_{\text{II}}$ spectrum.\cite{Ma04}

The simplest candidate that possesses the main features of the
D$_{\text{II}}$ center is the dicarbon antisite
(C$_{2}$)$_{\text{Si}}$ that we have already mentioned above in
connection with Fig.~\ref{fig:modes}. The dumbbell of two carbon atoms
sharing a silicon site (highlighted atoms in Fig.~\ref{fig:structure}
left) is surrounded by a tetrahedron of four carbon atoms. In 3C-SiC
the dumbbell is pointing in $\left< 100 \right>$ direction whereas in
4H-SiC it is oriented either along the $\left< 10\bar{1}1 \right>$ or
the $\left< \bar{1}011 \right>$ direction.  Depending on the Fermi
level, the defect may acquire the charge states from $2^{+}$ through
$2^{-}$.

We find that the dicarbon antisite has a relatively low formation
energy. For the neutral charge state the values are 6.7\,eV and
7.8\,eV in 3C and 4H polytypes, respectively. In 4H-SiC two possible
substitutional configurations exist, i.e.\ hexagonal (\emph{h}) and
cubic (\emph{k}), but the formation energy at these two sites only
differs by 0.1\,eV. These values are comparable to the formation
energies of C$_{\text{sp} \left< 100 \right>}$ and C$_{\text{sp\/Si}
  \left< 100 \right>}$.  Still, the calculated formation energy
indicates that its equilibrium abundance is too low for experimental
observation.  Hence kinetic effects should play an important role in
the dicarbon antisite formation. On the other hand, once created, the
defect has a very high stability, as reflected by the high binding
energy. Before we turn to kinetic aspects, we discuss the electronic
properties and the LVMs of the dicarbon antisite.

In 3C-SiC, the dicarbon antisite possesses two degenerate defect
orbitals with energy levels in the band gap. In the neutral charge
state these orbitals are occupied by two electrons, which can form
either a spin 0 (low-spin) or a spin 1 (high-spin) configuration.
According to the calculation, the high-spin state is about 80\,meV
lower in energy than the low-spin state.  This energy difference is,
however, too small to uniquely determine the ground state spin
configuration. For this reason we consider both options. The defect
wave function consists of bonds between the three upper carbon atoms
and the three lower carbon atoms of the defect tetrahedron (cf.
Fig.~\ref{fig:structure} left). In the low-spin configuration, one of
these bonds is fully occupied, resulting in a splitting of the defect
levels and a Jahn-Teller distortion, with the upper or the lower pair
of the tetrahedron being stretched. This reduces the original $D_{2d}$
symmetry to $C_{2v}$. In the high-spin configuration, both bonds are
equally occupied, retaining the degeneracy and the $D_{2d}$ symmetry.
As we discuss below, this geometrical difference is reflected in the
vibrational modes of the defect. In 4H-SiC, the high-spin state is
about 120\,meV lower than the low-spin state, which is---as in
3C-SiC---a very small energy difference. Due to the lower symmetry of
the crystal, the symmetry group is always $C_{1h}$. The defect's
structure on the cubic and the hexagonal site is qualitatively
similar.

\begin{table*}
  \caption{\label{tab:LVMs} LVMs of
    (C$_{2}$)$_{\text{Si}}$ and ((C$_{2}$)$_{\text{Si}}$)$_{2}$ in the
        neutral charge state in meV calculated at the LSDA lattice constant. 
        \emph{ls} denotes the low-spin, \emph{hs} the high-spin state. The
        subscripts $h$ and $k$ refer to the hexagonal and the cubic
        sites. For 4H-SiC and ((C$_{2}$)$_{\text{Si}}$)$_{2}$ the
        defect molecule approximation has been used. For the
        D$_{\text{II}}$ center only the five highlighted frequencies
        are given (cf.\ Refs.~\onlinecite{Pa73,Sr98_d2}).} 
  \begin{ruledtabular}
    \begin{tabular}{ccccccccccc}
      LVM & \multicolumn{4}{c}{3C} & \multicolumn{6}{c}{4H}\\
        \cline{2-5} \cline{6-11}
        & (C$_{2}$)$_{\text{Si}}$, ls & (C$_{2}$)$_{\text{Si}}$, hs &
        ((C$_{2}$)$_{\text{Si}}$)$_{2}$ & exp. &
        (C$_{2}$)$_{\text{Si,h}}$, ls & (C$_{2}$)$_{\text{Si,h}}$,hs &
        (C$_{2}$)$_{\text{Si,k}}$,ls & (C$_{2}$)$_{\text{Si,k}}$,hs
        & ((C$_{2}$)$_{\text{Si}}$)$_{2,\text{hk}}$ & exp.\\ \colrule
      1 & 116.4 & 107.6 & 123.3 & 127.8 & 103.0 & 100.2 & 102.3 & 101.5 & 119.6 & 127.1 \\
      2 & 128.5 & 123.2 & 125.6 & 130.9 & 121.9 & 121.0 & 119.7 & 120.7 & 132.2 & 129.8 \\
      3 & 139.4 & 141.7 & 127.1 & 146.3 & 136.2 & 136.4 & 135.0 & 136.2 & 147.0 & 146.1 \\
      4 & 148.3 & 141.8 & 139.3 & 152.2 & 139.3 & 137.6 & 139.1 & 137.6 & 160.2 & 152.4 \\
      5 & 177.5 & 177.5 & 168.6 & 164.7 & 179.4 & 177.1 & 178.0 & 177.0 & 162.4 & 164.4 \\
      6 &       &       & 169.7 &       &       &       &       &       &       &      \\

    \end{tabular}                                                    
  \end{ruledtabular}
\end{table*}

As outlined above, the dicarbon antisite is the simplest defect that
possesses LVMs that are similar to the phonon spectrum of the
D$_{\text{II}}$ center. The calculated frequencies (using a 64-sites
supercell) for the neutral charge state in low-spin and high-spin
configurations are listed in Tab.~\ref{tab:LVMs} along with the
experimental data for the five characteristic D$_{\text{II}}$ phonon
modes and our calculated results for the two neighboring dicarbon
antisites.  We see that in the 3C polytype the low spin configuration
provides an LVM pattern similar to the five phonon replicas of
D$_{\text{II}}$.  Similar results are obtained for 4H-SiC, except that
the frequencies of the modes 1 to 4 are lowered due to the defect
molecule approximation.

In spite of a different crystal symmetry, the vibrational patterns of
the LVMs in 3C- and 4H-SiC are very similar, owing to the localized
nature of the vibrational modes. The highest mode 5 is a stretching
vibration, where the two atoms of the dumbbell oscillate against each
other. Its calculated frequency is about 8\% higher than the
experimental value. However, this is the usual accuracy of these
defect calculations.\cite{Ca98} The modes 3 and 4 represent an
oscillation of one of the dumbbell atoms against its neighbors. In
3C-SiC, these modes are affected by the Jahn-Teller distortion of the
low-spin configuration. As expected from the symmetry, the modes are
degenerate in the high-spin and non-degenerate in the low-spin case.
This effect can also be seen in Fig.~\ref{fig:modes}, where the 216
sites calculation, although performed spin-unpolarized, has been
performed in a high-spin like configuration. Considering the small
energy difference between the high-spin and the low-spin configuration
both options should be allowed. In addition to the already mentioned
clustering of carbon atoms, this may be a source of further lines in
the luminescence spectra. In 4H-SiC, the modes 3 and 4 are always
non-degenerate, although the splitting in the high-spin configuration
is smaller than in the low-spin configuration, due to the lower
symmetry of the crystal field. The mode 2 describes the vibration of
the dumbbell along its axis against the enclosing tetrahedron. The
mode 1 is the least localized mode and therefore it is most affected
by the employed supercell technique. This mode corresponds to a
breathing-like vibration of the enclosing tetrahedron. While for the
low-spin case the breathing mode lies above the highest bulk
vibration, it is in resonance with the bulk spectrum and lies 8\,meV
below the calculated value of the highest bulk mode for the high-spin
case in a 64 sites cell.  This is in agreement with Gali~\emph{et al.}
(Ref.~\onlinecite{Ga02}) who considered the high-spin state of
(C$_{2}$)$_{\text{Si}}$.  In 4H-SiC, a similar behavior is not obvious
from our results, as the effect may be masked by the defect molecule
approximation. In addition to the high-frequency LVMs discussed above,
we also find maxima of the vibrational density of states in the phonon
band gap.

It is noteworthy that the results do not change significantly for
different possible charge states of the defect.\cite{Ma01b,Ma02} We
observe a slight lowering of the LVM's frequencies with the increase
of the number of localized electrons, resulting from an expansion of
the defect molecule due to the occupation of the bonds. For 3C-SiC,
the modes 3 and 4 are degenerate in the charge states $2^{+}$ and
$2^{-}$, since the electronic defect levels are either empty or fully
occupied, and a Jahn-Teller distortion is suppressed.

\section{Clustering of carbon atoms}
\label{sec:clustering}

Usually, the D$_{\text{II}}$ center is observed in ion-implanted and
annealed samples. It is stable up to temperatures of 1700\textcelsius.
This and the previously discussed low equilibrium concentration
suggest a kinetically controlled formation of this defect from
non-equilibrium vacancies and interstitials. At first, a carbon
antisite is formed via a recombination of a carbon interstitial with a
silicon vacancy. It merges then with a mobile carbon
split-interstitial, producing a dicarbon antisite with a high binding
energy that ranges from 3.9\,eV ((C$_{2}$)$_{\text{Si}}^{2+}$) to
5.1\,eV ((C$_{2}$)$_{\text{Si}}^{2-}$). The dicarbon antisite can
serve as a condensation center for larger clusters. Such clusters grow
by aggregating carbon split-interstitials around the core. Adding a
further carbon atom to the dicarbon antisite releases a binding
energy of about 4.8\,eV. Binding one more carbon interstitial to this
complex yields a very small cluster of four carbon atoms at a silicon
site (cf.  Fig.~\ref{fig:structure} right).  According to our
calculations the energy of 2.8\,eV is needed to remove a carbon atom
from this structure. However, for this defect we obtain LVMs well
above the D$_{\text{II}}$ spectrum. A detailed discussion of carbon
clusters will be given elsewhere.\cite{Ma04}

\begin{figure}
  \includegraphics[width=0.8\linewidth]{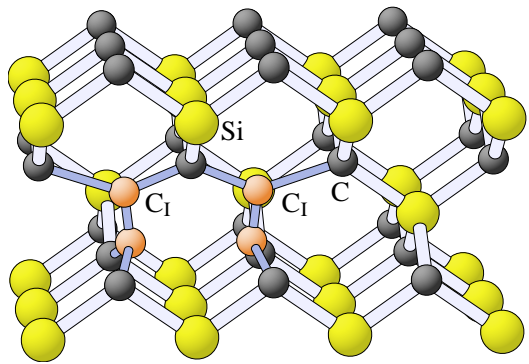}
  \caption{\label{fig:pair}
    Structure of a pair of dicarbon antisites
    ((C$_{2}$)$_{\text{Si}}$)$_{2}$ in 3C-SiC.}
\end{figure}

In addition to the clusters located around a single silicon site,
aggregates of several dicarbon antisites are also possible. A pair of
the dicarbon antisites in 3C-SiC ((C$_{2}$)$_{\text{Si}}$)$_{2}$ is
shown in Fig.~\ref{fig:pair}. In 4H-SiC three different
configurations, which can be compared to its 3C-SiC counterpart, are
possible when the two di-antisites occupy two cubic (\emph{kk}), two
hexagonal (\emph{hh}) or cubic-hexagonal (\emph{kh}) sites. As an
exemplary structure we have considered in detail the
\emph{kh}-configuration, which has a local geometry similar to the
3C-SiC case (Fig.~\ref{fig:pair}), but the bond length between the
neighboring dicarbon antisites is shorter.  We found that this defect
is very stable in both polytypes: in 3C-SiC it costs about 5.9\,eV to
remove a carbon atom, whereas in 4H-SiC this energy is 6.7\,eV. The
calculated LVMs of both complexes are given in Tab.~\ref{tab:LVMs}.
They have been calculated using the defect molecule approximation that
allows the vibration of the two dumbbells and of the surrounding
carbon neighbors within the 216 (3C-SiC) and 128 (4H-SiC) sites cell.
As discussed above, the numerical results are affected by the defect
molecule approximation.  Alternatively, one could work with the
smaller 64 sites cell which would allow to calculate the dynamical
matrix of the whole cell.  However, we find that in this cell the
electronic structure is so strongly distorted by the proximity effects
that it is unacceptable for the analysis of the LVMs. For the pair of
dicarbon antisites the highest almost degenerate modes 5 and 6
involve a stretching vibration of the dumbbells, similar to the lone
dicarbon antisite. Due to the interaction of the dumbbells these
modes become softer and combine into anti-symmetric and symmetric
vibrations.  The modes 3 and 4 represent a motion of the two lower and
upper carbon atoms of the dumbbells in bond direction against their
surrounding neighbors, respectively. In 4H-SiC, the defect shows a
similar behavior, but the energy of all modes is by about 8\,meV
lower than in 3C-SiC so that the lowest mode (with an energy of
114.4\,meV) drops into the bulk phonon spectrum.  This effect and also
the larger frequency difference between the stretching modes reflects
the stronger relaxation of the defect in this polytype. We have
recently learned that A. Gali \emph{et al.} calculated the LVMs of the
$kk$-configuration, finding results similar to the 3C-SiC values, and
speculated that this defect was the D$_{\text{II}}$
center.\cite{Ga03a} As the $kk$-configuration has a 3C-like
environment (in contrast to the $kh$-configuration), the similarity to
3C-SiC is quite natural. However, the conjecture that the
((C$_{2}$)$_{\text{Si}}$)$_{2}$ model completely describes the
D$_{\text{II}}$ center does not seem to be supported by experimental
data. Recent experiments have investigated the multitude of zero
phonon lines (ZPLs)\cite{Sr01} as well as the phonon
replicas\cite{Ca03} of the D$_{\text{II}}$ center. In 6H-SiC, Sridhara
\emph{et al.}\cite{Sr01} concluded from temperature and
stress-dependence experiments that the four ZPLs of D$_{\text{II}}$
stem from the different excitonic states of the bound exciton (one
ground state and three excited states), but not from different binding
centers of the same exciton. Regarding the phonon replicas in 4H-SiC
more than 12 LVMs and a single ZPL have been reported by Carlsson
\emph{et al.},\cite{Ca03} which is consistent with the earlier results
for 4H and 6H-SiC of Sridhara \emph{et al}.\cite{Sr98_d2} In addition,
as we noted above, a comparison of the spectra in 3C,\cite{Pa73} 4H
and 6H-SiC\cite{Sr98_d2} indicates that more than five LVMs may be
present also in 3C-SiC. The apparent large number of phonon replicas
in 4H and 6H-SiC makes it tempting to assume that they originate from
different geometrical configurations of the same simple defect. For
example, for the ((C$_{2}$)$_{\text{Si}}$)$_{2}$ in 4H-SiC this would
be $kk$, $hh$ and $kh$.  This assumption has, however, the following
implications.  Firstly, it follows that a larger number of LVMs should
be present in 6H-SiC than in 4H or 3C-SiC, since less or no
inequivalent sites are available in the two latter polytypes.
Secondly, the different configurations should give rise to a multitude
of exciton ground-state ZPLs. According to the conclusion of Sridhara
\emph{et al.}\cite{Sr01} and the rich vibrational structure contained
in all polytypes (cf.\ discussion above) this is not evident from the
experiments. This applies, in particular, to the
((C$_{2}$)$_{\text{Si}}$)$_{2}$ defect.  Furthermore, the LVM pattern
of this defect is, in fact, not compatible with the D$_{\text{II}}$
spectrum. The small splitting of the two highest LVMs (see
Tab.~\ref{tab:LVMs}), which persists through different polytypes and
configurational models, is not observed in the D$_{\text{II}}$
spectrum which has nearly equidistant lines. Thus the
((C$_{2}$)$_{\text{Si}}$)$_{2}$ model alone cannot explain all
features of the D$_{\text{II}}$ center.  Larger carbon clusters are
likely to be involved. For such clusters the differences between the
polytypes lose their importance. This may be responsible for the
polytype-independent LVMs of the D$_{\text{II}}$ centers. Further
experimental data are necessary to clarify the situation, especially
for the 3C polytype, which lacks inequivalent lattice sites.

\section{Conclusion}
\label{sec:conclusion}

In conclusion, we have analyzed the structural, electronic and
vibrational properties of the dicarbon antisite complex. The defect
possesses a characteristic vibrational pattern in the frequency range
of the D$_{\text{II}}$ spectrum, but cannot explain all the phonon
replicas observed. Yet, we find that the dicarbon antisite can act as
an aggregation center for carbon clusters that allow for a much richer
phonon spectrum. All the defect complexes presented above possess very
high dissociation energies, which guarantee their high thermal
stability. Due to the high formation energies, these defects can exist
in as-grown material only in very low concentrations but can be
kinetically created during the annealing. Yet, none of the considered
defect complexes can explain all features of the D$_{\text{II}}$
center. Most probably the D$_{\text{II}}$-type centers are related to
larger carbon clusters with the dicarbon antisite playing the role of
an elementary building block of the aggregates.

\begin{acknowledgements}
We acknowledge fruitful discussions with W.J.\ Choyke. This work has
been supported by the Deutsche Forschungsgemeinschaft within the
SiC Research Group.
\end{acknowledgements}

\end{document}